\documentclass[pra,floatfix,amsmath,twocolumn]{revtex4}
\usepackage{amssymb}
\usepackage{graphicx}
\usepackage{graphics}
\usepackage{amsmath}
\usepackage{amsthm}

\newcommand{\bea}{\begin{eqnarray}}
\newcommand{\eea}{\end{eqnarray}}
\def\bi{\begin{itemize}}
\def\ei{\end{itemize}}
\def\bc{\begin{center}}
\def\ec{\end{center}}

\def\C{\hbox{$\mit I$\kern-.7em$\mit C$}}
\def\R{\hbox{$\mit I$\kern-.6em$\mit R$}}

\def\ket#1{|#1\rangle}
\newcommand{\one}{\mbox{$1 \hspace{-1.0mm}  {\bf l}$}}
\def\tr{\mathrm{tr}}
\def\ket#1{\left| #1\right>}
\def\bra#1{\left< #1\right|}

\newcommand{\proj}[1]{\ket{#1}\bra{#1}}

\newtheorem{theorem}{Theorem}

\begin{document}

\title{Preparation of Entangled States by Quantum Markov Processes}

\author{B. Kraus$^{1}$, H.P. B\"uchler$^{2}$, S. Diehl$^{1,3}$, A. Kantian$^{1,3}$, A. Micheli$^{1,3}$, and P. Zoller$^{1,3}$}

\affiliation{$1$ Institute for Theoretical Physics, University of Innsbruck, Austria\\
$2$ Institute for Theoretical Physics III, University of Stuttgart,
Pfaffenwaldring 57, 70550 Stuttgart, Germany\\
$3$ Institute of Quantum Optics and Quantum Information of the
Austrian Academy of Sciences, Innsbruck, Austria }

\begin{abstract}

We investigate the possibility of using a dissipative process to
prepare a quantum system in a desired state. We derive for any
multipartite pure state a dissipative process for which this state
is the unique stationary state and solve the corresponding master
equation analytically. For certain states, like the Cluster states,
we use this process to show that the jump operators can be chosen
quasi--locally, i.e. they act non--trivially only on a few,
neighboring qubits. Furthermore, the relaxation time of this
dissipative process is independent of the number of subsystems. We
demonstrate the general formalism by considering arbitrary
MPS--PEPS states. In particular, we show that the ground state of
the AKLT--model can be prepared employing a quasi--local
dissipative process.

\end{abstract}
\maketitle

\section{Introduction}

Preparation of entangled pure quantum states is of interest in the
context of both quantum information and condensed matter physics.
In quantum information entangled states of qubits can act as a
resource for quantum computing, e.g. Cluster states \cite{Cluster}
in measurement based quantum computing \cite{RaBr01}, while in
condensed matter physics entangled states represent ground states
of strongly correlated systems.

A possible scenario for the preparation of entangled states of
interest is cooling the system to the ground state of an
appropriate many body Hamiltonian. Alternatively, we can generate a
state of interest from an initial pure state, e.g. a product state,
which can easily be prepared with available resources. This is
achieved either by coherent evolution generated by a system
Hamiltonian (i.e. a sequence of quantum gates), or, more generally,
by applying the most general physical transformation, which is
mathematically represented by a completely positive map. We will
discuss here another route: the preparation of entangled states by
designing dissipative processes, so that we drive the system via
non-equilibrium dynamics to a pure entangled state of interest
$\ket{\Psi}$ for long times, for any initial mixed state
represented by a system density operator $\rho$, i.e.
\begin{equation}
\rho\xrightarrow{t\rightarrow\infty}\ket{\Psi}\bra{\Psi}.\label{eq:goal}\end{equation}
In particular, we will consider a situation where the time
evolution of the system coupled to a reservoir can be described as
a Quantum Markov Process with dynamics obeying a master equation,
\begin{eqnarray}
\dot{\rho} & = & {\cal L}(\rho)\label{eq:masterequation}\\
 & \equiv & -i[H,\rho]+\sum_{\ell}g_{\ell}\left(2c_{\ell}\rho c_{\ell}^{\dagger}-c_{\ell}^{\dagger}c_{\ell}\rho-\rho c_{\ell}^{\dagger}c_{\ell}\right).\nonumber \end{eqnarray}
Here, $H$ represents a system Hamiltonian while the Liouvillian,
$L(\rho)=\sum_{\ell}g_{\ell}\left(2c_{\ell}\rho
c_{\ell}^{\dagger}-c_{\ell}^{\dagger}c_{\ell}\rho-\rho
c_{\ell}^{\dagger}c_{\ell}\right)$, can always be written in
Lindblad form with $c_{\ell}$ a set of {}``quantum jump operators''
and dissipation rates $g_{\ell}\geq 0$ \cite{Li76}. Such a
description in terms of a master equation is valid provided the
system dynamics is slow on the time scale of the reservoir
correlation time, as is the case for typical quantum optical
systems.

Thus the goal of \emph{dissipative entangled state preparation} is
to design quantum reservoirs and system-reservoir couplings, and to
identify necessary and sufficient conditions for the master
equation (\ref{eq:masterequation}), such that the desired pure
state of a many body system is obtained as the unique stationary
state. Indeed, it will be shown below that for any given
$\ket{\Psi}$ there is a master equation which yields the required
state as the unique pure state within a relaxation time
$T_{\textrm{relax}}\sim1/\min g_{k}$, independent of the number of
qubits.

We will be particularly interested in a situation where qubits (or
spin-1/2 particles) reside on a lattice. Thus it is natural to
restrict dissipation represented by the Liouvillian $L$ to
\emph{quasi-local} jump operators acting only on a neighborhood of
a given qubit, which raises the question of the class of states
which can be prepared with these resources. We will show below that
examples of states which can be generated include stabilizer
states, matrix-product states (MPS) or projected entangled pair
states (PEPS).

The results of the present paper are also of immediate relevance
for a non-equilibrium condensed matter physics where pure many body
states and quantum phases are prepared as a result of a driven
dissipative system dynamics. In a standard equilibrium situation of
condensed matter and cold atom physics, states close to the ground
state of the Hamiltonian, $H\ket{G}=E_{G}\ket{G}$, are prepared by
cooling the system, where in particular for a finite system
$\rho\sim
e^{-H/k_{\mathrm{B}}T}\rightarrow\ket{\textrm{G}}\bra{\textrm{G}}$
for temperature $T\rightarrow0$. This has already lead to the
preparation of intriguing quantum phases
\cite{greiner02,stoferle04,kinoshita04,paredes04,folling07,zwierlein07,Bloch07}.
In contrast, we obtain a pure state representing a non-equilibrium
quantum phase as a result of the dynamics (\ref{eq:goal}) with the
master equation (\ref{eq:masterequation}). In recent work we have
provided examples of master equations, including the example of
non-interacting bosons and paired interacting fermions
corresponding to cold atoms moving in an optical lattice which are
driven by coupling to quasi--local reservoirs into pure states
exhibiting long range order. While the goal of Ref. \cite{BEC} was
to study non-equilibrium condensates, Luttinger liquids and
Kosterlitz-Thouless phases, in the present work we will provide the
uniqueness proofs for the corresponding driven dissipative
dynamics. Furthermore, the results of the present paper are of
direct relevance for non-equilibrium spin models. As an example we
will discuss a master equation whose unique steady state is the
ground state of the familiar AKLT model \cite{Aklt}.

The outline of the paper is as follows. First of all, we summarize
some properties of the master equation. In Section III we
characterize all stationary pure states. Since we are interested in
unique stationary states, we derive a sufficient condition for the
uniqueness of the stationary state. In Section IV we derive, for
any multipartite state, a dissipative process which can be used to
prepare this state. That is, we construct a dissipative process for
which the desired pure state is the unique stationary state. For
this process it is simple to solve the underlying master equation
analytically and to show that the relaxation time of the system is
independent of the number of subsystems. In Section V we finally
show that for certain states, like the 2D-Cluster states
\cite{Cluster}, this construction can be used to choose jump
operators quasi--locally. We furthermore derive a quasi--local
dissipative process which has a general PEPS \cite{VeCi04} as the
unique stationary pure state. Also in the context of PEPS, we
consider the ground state of the familiar AKLT--model \cite{Aklt}
and derive the dissipative process for which this state is the
unique stationary state. We also prove the uniqueness of the driven
noninteracting BEC and the $\eta$-condensate of paired fermions
given specific dissipative processes, complementing the work done
on these states in~\cite{BEC}.

\section{Preliminaries and Notation}

We are interested in the stationary solutions of the master
equation $\dot{\rho}={\cal L}(\rho)$, presented in Eq.
(\ref{eq:masterequation}). We write ${\cal L}(\rho)$ as ${\cal
L}(\rho)={\cal E}(\rho)-Q^\dagger\rho-\rho Q,$ where ${\cal
E}(\rho)=2\sum_{l} g_l 2 c_l\rho c_l^\dagger$ is a completely
positive map and $Q=P-iH$, with $P=\sum_{l} g_l c_l^\dagger c_l$ a
positive semidefinite operator. Sometimes we denote ${\cal L}$, as
given in Eq. (\ref{eq:masterequation}), by ${\cal L}_{\{H, c_l\}}$
and by ${\cal L}_{\{ c_l\}}$ if we consider a purely dissipative
process. Note that the partition in the Hamiltonian and the
dissipative part is unique if the operators $c_l$ are traceless and
orthonormal \cite{WoCi06}.

Since the master equation is linear, the eigenvalue equation ${\cal
L}(\sigma)=\lambda \sigma$ can be written as a matrix equation. Due
to the fact that $\tr({\cal L}(\sigma))=0$ for any $\sigma$, the
eigenvectors to eigenvalues different than zero must be traceless.
The eigenvalues can be complex, however, the real part of the
eigenvalues is not positive (see for instance \cite{BaNa08}).
Considering a purely dissipative process, with hermitian quantum
jump operators $c_l$, all the eigenvalues are real. This is due to
the fact that in this case the matrix corresponding to the
Liouvillian is hermitian. However, in general the matrix is not
hermitian, it is not even diagonalizable and has therefore
generalized eigenvectors, $\sigma$ \cite{Le87}. Furthermore, since
${\cal L}(\sigma^\dagger)=({\cal L}(\sigma))^\dagger$ the
eigenvalues occur in pairs of the form $\lambda, \lambda^\ast$ with
the corresponding eigenvectors $\sigma$ and $\sigma^\dagger$. The
set of all proper and all generalized eigenvectors, $\{\sigma_i\}$,
forms a basis in the operator space. Expanding $\rho(0)$ in this
basis, $\rho(0)=\sum_i c_i \sigma_i$ we obtain $\rho(t)= \sum_i
\sum_j c^j_i(t)\sigma_j e^{\lambda_i t}$ where $c^j_i(t)$ are
polynomials of degree less than the largest order of the Jordan
block corresponding to eigenvalue $\lambda_i$ \cite{Le87}.

We are interested in the stationary states of the evolution. That
is we want to find the states, $\rho$, for which ${\cal L}_{\{H,
c_l\}}(\rho)=0$. In order to do so, we use the following notation.
By $K(X)$ [$R(X)$] we denote the kernel [range] of a hermitian
operator $X$ and $r(X)$ denotes the rank of $X$. Note that the pure
states, which are not affected at all by the dissipative process
coincide with the kernel of $P$, $K(P)=\{\ket{\Psi} \mbox{ such
that } c_l \ket{\Psi}=0 \forall l \in {\bf L}\}$. Thus, if
$\ket{\Psi}\in D_{\{H, c_l\}} \equiv K(P)\bigcap ES(H)$, where
$ES(H)$ denotes the eigenspace of $H$, then ${\cal L}_{\{H,
c_l\}}(\proj{\Psi})=0$. This implies that any state $\rho$ with
$R(\rho)=\mbox{span} \{\ket{\phi_i}\}$ such that
$\{\ket{\phi_i}\}\subseteq D_{\{H, c_l\}}$ is a stationary state.
We call these states dark states. Whenever it is clear from the
context we omit the subscripts ${\{H, c_l\}}$ and write for
instance simply $D$ for the subset of dark states.

\section{Pure Stationary States and Uniqueness of Stationary States}

In this section we first of all characterize all pure stationary
states. We show that a pure state is a stationary state of some
dissipative process iff it is a dark state of some other. Then we
derive a sufficient condition for the uniqueness of stationary
states.

%%%%%%%%%%%%%%%%%%%%%%%%%%%%%%%%%%%%%%%%%%%%%%%%%%%%%%%%%%%%%%%%%%%%%%%%
Let us first of all, consider a given dissipative process and
characterize the pure states, which are stationary states of it.
\begin{theorem}\label{Thpure}
Let ${\cal L}$ be defined as in Eq. (\ref{eq:masterequation}). Then
${\cal L}(\proj{\Phi})=0$ iff the following two conditions are
fulfilled: \bi \item[(1)] $\ket{\chi}\equiv Q^\dagger
\ket{\Phi}=\lambda\ket{\Phi}$ for some $\lambda\in \C$. \item[(2)]
$ c_l\ket{\Phi}=\lambda_l \ket{\Phi} $ $\forall l\in {\bf L}$, for
some $\lambda_l\in \C$ with $\sum_l g_l |\lambda_l|^2=Re(\lambda)$,
where $Re(x)$ denotes the real part of $x$.\ei\end{theorem}

\begin{proof} ${\cal L}(\proj{\Phi})=0$ iff \bea
\label{eq_AB} 2\sum_l g_l \proj{\Psi_l}={\cal
E}(\proj{\Phi})=\ket{\Phi}\bra{\chi}+\ket{\chi}\bra{\Phi},\eea
where $\ket{\Psi_l}=c_l\ket{\Phi}$. Therefore, the operator
$A\equiv \ket{\Phi}\bra{\chi}+\ket{\chi}\bra{\Phi}$ must be
positive semidefinite. It can be easily verified that $A\geq 0$ iff
$r(A)\leq 1$. Thus, $\ket{\chi}=\lambda \ket{\Phi}$ for some
$\lambda\in \C$ and $A=2Re(\lambda)\proj{\Phi}$. The fact that all
$\ket{\Psi_l}$ are in the range of ${\cal E}(\proj{\Phi})$ implies
then, that Eq. (\ref{eq_AB}) is fulfilled iff
$\ket{\Psi_l}=\lambda_l\ket{\Phi}$ with $\sum_l g_l
|\lambda_l|^2=Re(\lambda)$.

\end{proof}

Using this theorem it is now easy to characterize all pure states
for which there exists a Liouvillian such that the state is a
stationary state of the master equation. Defining $c_l^\prime=
c_l-\lambda_l \one$ and $H^\prime =H-i\sum_l g_l \lambda_l
(c_l^\prime)^\dagger +i \sum_l g_l \lambda_l^\ast c_l^\prime$ it is
straightforward to show that the conditions above imply that 1)$
H^\prime \ket{\Phi}=\lambda \ket{\Phi}$, with $\lambda\in \R$ and
2)$ c_l^\prime \ket{\Phi}=0$ $ \forall l\in \{1,\ldots m\}$. Thus,
we have that there exists a Liouvillian ${\cal L}$ such that ${\cal
L}(\proj{\Phi})=0$ iff there exists a set of operators
$\{c_l^\prime\}_{l=1,\ldots m}$, a Hamiltonian $H^\prime$ and some
$\lambda\in \R$, such that the following two conditions are
fulfilled: \bi \item[($1^\prime$)] $ c^\prime_l\ket{\Phi}=0 $
$\forall l\in {1,\ldots m}$ \item[($2^\prime$)] $
H^\prime\ket{\Phi}=\lambda \ket{\Phi}.$\ei Note that the two
conditions $(1^\prime)$ and $(2^\prime)$ are equivalent to
$\ket{\Phi}\in D_{\{H^\prime, c_l^\prime\}}$. Therefore,
$\ket{\Phi}$ is a stationary state, iff it is a dark state for some
other physical process. Thus, in order to design a dissipative
process, which leads to the desired pure state, we have to find a
set of operators (or a single operator) which have only one common
eigenstate, the corresponding eigenvalue can be chosen to be zero.
Due to the results presented above, we know that the corresponding
dissipative process will have the desired state as the unique pure
stationary state. However, since we want to use this process for
state preparation, we have to guarantee that there exists no mixed
stationary state. How this can be ensured will be shown next.

We consider the general master equation given in Eq.
(\ref{eq:masterequation}). Note that the corresponding set of dark
states is in general a set, not a subspace. However, since we want
to use the dissipative process for state preparation, or to drive
the system to a certain (higher--dimensional) subspace, we will
consider here the situation where $D$ is a subspace, i.e. all the
states in $D$ correspond to the same eigenvalue of the Hamiltonian.
We are going to show that if there exists a stationary state,
$\rho$, which is not a dark state, i.e. $R(\rho) \nsubseteq D$,
then there must exist a subspace of the Hilbert space, ${\cal H}$,
which is left invariant under the operators $\{c_l\}$.

\begin{theorem} \label{Thmixed} If there exists no subspace $S \subseteq {\cal H}$ with
$S \perp D$ such that $c_l S \subseteq S$ $\forall c_l$, then the
only stationary states are the dark states.\end{theorem}

\begin{proof} We prove the statement by contradiction. That is we assume that
there exists a state $\rho$, with $R(\rho)=S^\prime \nsubseteq D$
such that ${\cal L}(\rho)=0$ and show that this implies that there
exists a subspace $S \subseteq {\cal H}$ with $S \perp D$ such that
$c_l S \subseteq S$ $\forall c_l$. Using the notation of Eq.
(\ref{eq:masterequation}) and the fact that ${\cal E}$ is a
completely positive map, we have that ${\cal
E}(\rho)=Q^\dagger\rho+\rho Q \geq 0$. Since ${\cal E}(\rho)$ is a
positive semidefinite operator we have that $R({\cal
E}(\rho))=\mbox{span} \{c_l \ket{\xi}, l\in L, \ket{\xi}\in
S^\prime\}$. We are going to show now that $R({\cal E}(\rho))$ must
be within $S^\prime$ which implies that $c_l S^\prime \subseteq
S^\prime$ $\forall c_l$. Since $Q^\dagger \rho+\rho Q$ must be
positive semidefinite $K(Q^\dagger\rho+\rho Q)= \{\ket{\Phi} \mbox{
such that } \bra{\Phi}Q^\dagger\rho+\rho Q \ket{\Phi}=0\} \supseteq
K(\rho) \bigcup D$. Therefore we have that $K({\cal E}(\rho))
\supseteq K(\rho) \bigcup D$ which implies that $R({\cal E}(\rho))
\subseteq R(\rho) \bigcap D^{\perp}$, where $D^{\perp}$ denotes the
orthogonal complement of $D$. We have shown now that $\{c_l
\ket{\xi}, l\in {\bf L}, \ket{\xi}\in S^\prime\}\subseteq S^\prime
\bigcap D^{\perp}$. It remains to show that if such a $S^\prime
\nsubseteq D$ exits, then there exists a subspace $S\subseteq
D^\perp$ fulfilling that $c_l S \subseteq S$ $\forall c_l$. Note
that the set $D^\prime=\{\ket{\Psi} \mbox{ such that } Q^\dagger
\ket{\Psi}=i\lambda \ket{\Psi}, \mbox{ with }\lambda \in \R\}$ is
equal to $D$ \footnote{This can be seen as follows. It is obvious
that $D\subset D^\prime$ to show the inverse, we use that
$\ket{\Psi}\in D^\prime$ implies that $\bra{\Psi} P \ket{\Psi}+ i
\bra{\Psi} H \ket{\Psi}=i\lambda $. Since both, $P$ and $H$ are
hermitian, (and $\lambda\in \R$) this last equation can only be
fulfilled if $\ket{\Psi}\in D$.}. Using again that
$Q^\dagger\rho+\rho Q$ must be positive semidefinite which implies
that $\bra{\Phi}Q^\dagger\rho+\rho Q \ket{\Phi}=0$ iff
$(Q^\dagger\rho+\rho Q) \ket{\Phi}=0$ we have that $Q^\dagger\rho
\ket{\Phi}=i \lambda \rho \ket{\Phi}$ $\forall \ket{\Phi}\in D$.
Thus, $\rho \ket{\Phi}\in D$ $\forall \ket{\Phi}\in D$, which
implies that $\rho$ can be decomposed as $\rho=\rho_D+
\rho_{D^\perp}$, where $R(\rho_D)\subseteq D$ and
$R(\rho_{D^\perp})\subseteq D^\perp$. Now, since ${\cal
L}(\rho)={\cal L}(\rho_{D^\perp})$ we only have to chose
$S=R(\rho_{D^\perp})$.

\end{proof}

This proof shows that if there exists a stationary state, $\rho$
which is not a dark state, i.e. $R(\rho)\nsubseteq D$ then
$\rho=\rho_D+\rho_{D^\perp}$ with $R(\rho_D)\subseteq D$ and
$R(\rho_{D^\perp})\subseteq D^\perp$ such that $R(\rho_{D^\perp})$
is invariant under the operators $\{c_l\}$.

In the next sections we use the results presented above to design
dissipative processes which can be used for state preparation. That
is, we derive the jump operators, such that the system is driven
into the unique stationary state. Due to the results above this
state can be chosen to be a dark state of the process.

Note that this goal can never be achieved using hermitian jump
operators $c_l$. The reason for that is the following. If all
operators $c_l$ are hermitian, then the master equation can be
written as \bea {\cal L}(\rho)=\sum -i[H,\rho]+g_l [[c_l,\rho],
c_l].\eea Thus, any operator which commutes with $\{c_l\}$ and $H$
is a stationary state, for instance, the completely mixed state
$\propto \one$ is stationary.

%%%%%%%%%%%%%%%%%%%%%%%%%%%%%%%%%%%%%%%%%%%%%%%%%%%%%%%%%%%%%%%%%%%%%

\section{Construction of dissipative processes leading to the
desired state}

In this section we show that one can design the system--reservoir
coupling in such a way that any multipartite state can be obtained
as the unique stationary state of a dissipative process.
Furthermore, we solve the corresponding master equation
analytically by deriving the whole spectrum of the corresponding
Liouvillian. Given this solution it is then easy to compute, any
relevant quantity of the process, like for instance the relaxation
time, or any correlation function. In fact, we show that the
relaxation time is independent of the number of subsystems. Even
though these facts are not very surprising, we will use the
constructed process to show that for certain states the jump
operators can be chosen quasi--locally. In general such a
construction will not lead to a quasi--local dissipative process.
Therefore, we demonstrate how the quasi--local operators can be
constructed given a quasi--local description of the state at the
end of this section. We start out by considering $n$--qubit states
and generalize later the formalism to a $d$--level system.

For any $n$--qubit state $\ket{\Psi}=U\ket{0}^{\otimes n}$ we
construct a set of operators $\{c_k\}_{k=1}^n$ such that the unique
stationary state of the dissipative process, described by
$\dot{\rho}={\cal L}_{\{c_k\}}(\rho)$, is $\ket{\Psi}$. Note that
$\ket{0}^{\otimes n}\equiv \ket{0^n}$ is the unique stationary
state of the dissipative process with jump operators
$d_k=\sigma_{-}^{(k)}=\one_1\otimes. \ldots \otimes
\ket{0}_k\bra{1}\otimes \one_n$, i.e. $d_k$ acts non--trivially
only on the $k$--th qubit. This can be easily seen, since
$d_k\ket{\phi}=0$ $\forall k$ iff $\ket{\phi}=\ket{0^n}$, which
implies that $\ket{0^n}$ is the unique pure stationary state (see
Theorem \ref{Thpure}). Furthermore, it is straightforward to show
that for any $\ket{\chi}\perp \ket{0^n}$ there exists a monomial of
the jump operators, $P(\ket{\chi})$ such that
$\bra{0^n}P(\ket{\chi})\ket{\chi}\neq 0$. This shows that
$\ket{0^n}$ is the unique stationary state of ${\cal L}_{\{d_k\}}$.

We construct now the operators $c_k$ which lead to the unique
stationary state $\ket{\Psi}=U\ket{0^n}$. Defining $c_k=U d_k
U^\dagger$ we have $c_k\ket{\phi}=0$ iff $\ket{\phi}=U \ket{0^
n}=\ket{\Psi}$. One can use the same arguments as above to show
that this is the unique stationary state. This immediately implies
that if the jump operators $c_k$ can be written as
$U(\one_{1,\ldots k-1} \otimes \ket{0}_k\bra{1} \otimes
\one_{k+1,n})U^\dagger$ and $\ket{\Psi}\equiv U\ket{0^n}$ is an
eigenstate of the Hamiltonian $H$, then $\ket{\Psi}$ is the unique
stationary state of ${\cal L}_{\{H,c_k\}}$.

Note that the statements above would not be changed if we would use
an invertible matrix, $X$, instead of a unitary, as long as this
does not lead to more common eigenstates of the operators $\{X c_k
X^{-1}\}$.

\subsection{Analytic Solution of the Master Equation}

In this subsection we solve analytically the master equation
$\dot{\rho}={\cal L}_{\{c_k\}}(\rho)$, for any set of operators
$\{c_k\}$ which are unitarily equivalent to the set
$\{\sigma_{-}^{(k)}\}$, i.e. $c_k=U\sigma_{-}^{(k)}U^\dagger$ for
some unitary $U$.

In order to compute the eigenvalues and eigenvectors of the
Liouvillian ${\cal L}_{\{c_k\}}$ we note that ${\cal
L}_{\{U\sigma_{-}^{(k)} U^\dagger\}}(\sigma)=U{\cal
L}_{\{\sigma_{-}^{(k)}\}}(U^\dagger \sigma U) U^\dagger$.
Therefore, ${\cal L}_{\{U\sigma_{-}^{(k)}
U^\dagger\}}(\sigma)=\lambda \sigma$ iff ${\cal
L}_{\{\sigma_{-}^{(k)}\}}(U^\dagger \sigma U)= \lambda U^\dagger
\sigma U $. Thus, computing the eigenvectors and eigenvalues of
${\cal L}_{\{\sigma_{-}^{(k)}\}}$ gives us immediately the
eigenvalues and eigenvectors of ${\cal L}_{\{c_k\}}$.

Since ${\cal L}_{\{\sigma_{-}^{(k)}\}}$ describes the situation
where $n$ qubits are interacting with identical and independent
reservoirs, we only have to find the eigenvectors and eigenvalues
of ${\cal L}_{\{\sigma_{-}^{(k)}\}}$ for one fixed $k$. This
describes the situation where a single two--level system is
interacting with a thermal bath. The solution to the eigenvalue
problem is well--known \cite{GardinerZoller00,Barnett}. We find
${\cal L}_{\{\sigma_{-}^{(k)}\}}(\sigma_i^{(k)})=\lambda_i
\sigma_i^{(k)}$ iff $\sigma_i^{(k)}\in
\Sigma_k\equiv\{\sigma_0^{(k)}=\proj{0},\sigma_3^{(k)}=\sigma_z^{(k)}\}\bigcup
\mbox{span}\{\sigma_1^{(k)}=\sigma_{x}^{(k)},\sigma_2^{(k)}=\sigma_{y}^{(k)}\}$
and the corresponding eigenvalues are $\lambda_0=0,\lambda_3=-2
g_k,$ and the two--fold degenerate eigenvalue is
$\lambda_1=\lambda_2=-g_k$. Thus, a basis of the eigenvectors of
the total Liouvillian is $\{\sigma_{i_1}^{(1)}\otimes \ldots
\sigma_{i_n}^{(n)} \mbox{ with } \sigma_{i_k}^{(k)} \in \Sigma_k
\}_{i_k\in \{0,1,2,3\}}$ and the corresponding eigenvalues are
$\lambda_{i_1,\ldots, i_n}\equiv \sum_{k} \lambda_{i_k}$. Note that
the eigenvectors of ${\cal L}_{\{c_k\}}$ are just the unitary
transformation of the eigenvectors of ${\cal
L}_{\{\sigma_{-}^{(k)}\}}$ with the same eigenvalues. Thus, writing
the initial state in the eigenbasis, $\rho(0)=U\sum_{i_1,\ldots
i_n} c_{i_1,\ldots i_n} \sigma_{i_1}^{(1)}\otimes \ldots
\sigma_{i_n}^{(n)}U^\dagger$ we obtain \bea \rho(t)&=&e^{{\cal L}
t}(\rho(0))\\ \nonumber &=&U\sum_{i_1,\ldots i_n} c_{i_1,\ldots
i_n} e^{-\lambda_{i_1,\ldots, i_n}t} \sigma_{i_1}^{(1)}\otimes
\ldots \sigma_{i_n}^{(n)}U^\dagger.\eea

Let us stress here the fact that there exists no purely imaginary
eigenvalue. If such a pair (recall that complex eigenvalues occur
in pairs, see introduction) existed then the additional condition
that $\lim_{t\rightarrow \infty} \rho(t)=\rho_{ss}$, where
$\rho_{ss}$ denotes the stationary state, would not be satisfied.
In order to see how fast the system is driven into the stationary
state, we compute the relaxation time, $T_{\textrm{relax}}$ which
is defined as the inverse of the minimum of the negative real part
of $\lambda_{i_1,\ldots, i_n}$ different than zero. We find
$T_{\textrm{relax}}=1/\mbox{min}_k g_k$ and is therefore given by
the minimal coupling constant. Note that it is not very surprising
that $T_{\textrm{relax}} \geq 1/\mbox{min}_k g_k$, since the system
cannot be driven faster into the stationary state. The reason why
this amount of time is already sufficient is because the evolution
of the qubits can be decoupled. In other words, for fixed coupling
constants, the relaxation time is constant in the number of qubits.

It is straightforward to generalize this formalism to $d$--level
systems. One simply has to replace the operators $\ket{0}\bra{1}$
by the a matrix $J_d
=\ket{0}\bra{1}+\ket{1}\bra{2}+\ket{2}\bra{3}+\ldots
\ket{d-1}\bra{d}$, i.e. a Jordan matrix with only one eigenstate
(here to eigenvalue zero)\footnote{Note that any factor in front of
the terms $\ket{j-1}{j}$ would not change the argument. Thus, $J_d$
is like the annihilation operator for finite dimensions.}. Only the
state $\ket{0}$ is a proper eigenstate of $J_d$. All the other
computational basis states are generalized eigenvectors, which
means that for any $k\neq 0$ there exists an $i$ such that $J_d^i
\ket{k}=\ket{0}$. This implies that for any state $\ket{\phi}$
there exists a polynomial of the operators $J_d$, $P(J_d)$, such
that the overlap $\bra {0} P(J_d)\ket{\phi}$ is not vanishing,
which is exactly the property that we need in order to prove that
the stationary state $\ket{0^n}$ or respectively $U\ket{0^n}$ is
the unique stationary state. Since also in this case, the master
equation decouples for the different subsystems it is
straightforward to solve it analytically.

\subsection{Quasi--local dissipative Processes}

If one wants to prepare the state $\ket{\Psi}$ using a dissipative
process, then one will be interested in a simple physical
interaction between the reservoir and the system. One requirement,
for instance, could be that the operators $c_k$ are quasi--local,
which means that they act non--trivially only on a small number of
qubits. Depending on the state $\ket{\Psi}$ the operators $c_k$
might be chosen quasi--local, as we are going to show next. Note
however, that this cannot be true for any state. The reason for
this is the following.

As shown before, for any $n$--qubit state $\ket{\Psi}$ one can find
$n$ operators $c_k$ which uniquely define the state $\ket{\Psi}$,
in the sense that $\ket{\Psi}$ is the only state which is a (right)
eigenstate to eigenvalue $0$ of all operators $c_k$. Thus, a
description of the state $\ket{\Psi}$ is the set of operators
$\{c_k\}_{k=1}^n$. In general one might need more than one operator
per subsystem. We denote the corresponding set of operators by
$\{c_k^\alpha\}$, where $\alpha=1,\ldots, d$, for some $d$, and
call this set a quasi--local description of $\ket{\Psi}$ if all
$c_k^\alpha$ are quasi--local. The Quantum Kolmogorov complexity,
i.e. the number of classical bits required to describe the state
$\ket{\Psi}$ equals the classical Kolmogorov Complexity of the set
$\{c_k^\alpha\}$. If all these operators would be quasi--local,
then the Kolmogorov Complexity scales only polynomially with the
number of qubits. It is known however, that for any $n$ there
exists a $n$--qubit state, whose Quantum Kolmogorov Complexity
scales exponentially with the number of qubits \cite{Mora}.

Before we discuss more generalized schemes we want to use the
process discussed above to show that certain states can be prepared
using quasi--local dissipative processes. We consider the state
$\ket{\Psi}=U\ket{0}^{\otimes n}$. Now, suppose that
$U=U_{1}U_{2}\ldots U_n$, where each of the unitaries $U_k$ commute
with each other and are quasi--local. Let us assume that $U_k$ is
acting on particles $k-1,k,k+1$. Then the jump operators $c_k=U
\sigma_k U^\dagger=U_{k-1}U_k U_{k+1} \sigma_k (U_{k-1}U_k
U_{k+1})^\dagger$ are also quasi--local and the dissipative process
corresponding to $c_k$ has as a unique stationary state the state
$\ket{\Psi}$. As shown before, for such a process the relaxation
time is constant, i.e. independent of the number of subsystems.

For certain cases such a simple construction will not be possible.
Therefore, we describe here a general method of deriving the jump
operators, i.e. the dissipative process, which give rise to the
desired state. Since the jump operators are not hermitian (at least
not all of them can be chosen to be hermitian), they might not be
diagonalizable. However, one can use the Jordan normal form to gain
some insight in the necessary properties of these operators
\cite{Horn}. The jump operators must be chosen such that there
exists only one common eigenstate to eigenvalue zero. The Jordan
decomposition of a matrix $c$ is \bea c= S J S^{-1}, \eea where $J$
denotes the Jordan matrix of $c$ and $S$ is an invertible matrix.
$J$ is a block diagonal matrix with $d_i\times d_i$ diagonal blocks
\bea
J_{d_i}(\lambda_i)=\begin{pmatrix} \lambda_i & 1 &  & & \\
0 & \lambda_i & 1 & & &  \\
& & \ddots & \ddots & \\
 & & & \lambda_i & 1\\ & & & & \lambda_i
\end{pmatrix}.\eea

The number of Jordan blocks is the number of linearly independent
eigenvectors of $c$. Let us for simplicity consider here the case
where $S$ is a unitary, $U$. The proper eigenvectors of $c$ are
then $U \ket{e_{f(k)}}$, where $f(k)=\sum_{i=0}^k d_i +1$ and
$\ket{e_k}$ denotes the standard basis. A simple example is the
operator $\sigma_{-}$. The only eigenstate is the state $\ket{0}$.
Considering multipartite entangled states, the operators $c_k$ must
have more than a single eigenstate to eigenvalue $0$ (like
$\one\otimes \sigma^k_-\otimes \one$ has). Therefore the matrices
$c_k$ will have Jordan matrices with several Jordan block and all
eigenvalues $0$. The unique eigenstate which is common to all
operators is the state one wants to prepare, $\ket{\Psi}$. Apart
from that, one has to assure that any other state can be mapped
into some state having non--vanishing overlap with $\ket{\Psi}$.

One might also chose a single jump operator, which has only one
proper eigenstate (which corresponds to eigenvalue zero) (see also
\cite{BaNa08}). In order to be more precise, let us denote by
$\{\ket{\phi_i}\}_{i=0}^{2^n-1}$ an orthonormal basis with
$\ket{\phi_0}=\ket{\Psi}$ being the state we want to prepare. The
operator $C=\sum_{i=0}^{2^n-1} \ket{\phi_i, }\bra{\phi_{i+1}}$ has
only one proper eigenstate, namely $\ket{\Psi}$. For the generalized
eigenvectors, $\ket{\phi_i}$, for $i>0$ it holds that $C^i
\ket{\phi_i}=\ket{\Psi}$. Thus, for any state $\ket{\chi}$, there
exists a $k$ such that $\bra{\Psi}C^k\ket{\chi}\neq 0$. This shows
that the single operator $C$ leads, as the operators $\{c^{(l)}\}$
described above, to the unique stationary state $\ket{\Psi}$.
However, in contrast to the operators $\{c_k\}$ the operator $C$
will act non--trivially on all subsystems and is therefore (in
general) harder to implement experimentally.

In summary, a sufficient condition for the existence of a
quasi--local dissipative process, which can be employed to prepare
a certain state is the following. First of all, the state must have
a quasi--local description, i.e. there exists a set of operators,
$\{c_k^\alpha\}$ such that the only common eigenstate is
$\ket{\Psi}$. This implies that there exists only one dark state
for the corresponding dissipative process. Second, if there exist
polynomials $P_i(\{c_k^\alpha\})$ such that the states
$P^\dagger_i(\{c_k^\alpha\})\ket{\Psi}$ form a basis in the Hilbert
space, then $\ket{\Psi}$ is the unique stationary state. More
generally, if there exist polynomials $P_i(\{c_k^\alpha\})$ such
that ${\cal E}(\proj{\Psi})=\sum_i
P^\dagger_i(\{c_k^\alpha\})\proj{\Psi}P_i(\{c_k^\alpha\})= X $
where $X$ is invertible, then $\ket{\Psi}$ is the unique stationary
state. This can be seen as follows. If ${\cal E}(\proj{\Psi})=X$,
with $X$ invertible, then we have for any state $\ket{\chi}\in
{\cal H}$, $\bra{\chi}{\cal E}(\proj{\Psi})\ket{\chi}\neq 0$. Since
${\cal E}(\proj{\Psi})$ is a sum of positive semi--definite
operators this implies that there exists at least one term,
$P^\dagger_i(\{c_k^\alpha\})\proj{\Psi}P_i(\{c_k^\alpha\})$, which
has a non--vanishing overlap with the state $\proj{\chi}$.

\section{Examples}

In this section we illustrate the formalism described above by
applying it to several examples. In subsections
\ref{subsec:drivenBEC} and \ref{subsec:drivenEta} we show that the
dissipative processes which we analyzed in detail in \cite{BEC},
have unique stationary states, namely the BEC--state and the
$\eta$-condensate respectively. In subsection
\ref{subsec:drivenStab} we derive the processes for stabilizer
states, and in subsection \ref{subsec:drivenMPS} we consider
arbitrary PEPS states and show how, for instance the ground state
of the AKLT--model can be generated with a quasi--local dissipative
process.

\subsection{Example: Driven Dissipative Hubbard Dynamics of Bosons
on a Lattice}\label{subsec:drivenBEC}

In Ref. \cite{BEC} we have described a \emph{driven dissipative
Hubbard dynamics }of bosonic particles on a lattice. The
corresponding dynamics was written in terms of a master equation
(\ref{eq:masterequation}) with $H$ a Hubbard Hamiltonian,
containing the coherent hopping of particles between the sites of
the lattice, and their interaction, and where a Liouvillian with
quasi-local jump operators was designed to drive the system into a
non-equilibrium condensate of bosons or paired fermions. While Ref.
\cite{BEC} focused on non-equilibrium condensed matter aspects and
in particular on the effect of interactions, and implementation of
this master equation with cold atoms, we claimed -- but did not
prove -- the uniqueness of the many body dark states for this
master equation. This missing proof will be provided here based on
the theorems derived in the previous sections.

We consider $N$ bosons hopping on a $d$-dimensional lattice with
Hamiltonian\begin{equation} H_B=H_{0}+V\equiv-J\sum_{\langle
i,j\rangle}a_{i}^{\dagger}a_{j}+\tfrac{1}{2}U\sum_{i}a_{i}^{\dagger2}a_{i}^{2},\label{eq:Hubbard}\end{equation}
which is written as the sum of a kinetic and interaction energy
term. Here $a_{i}$ denotes the destruction operator for a boson on
site $i$, and $\langle i,j\rangle$ denotes adjacent lattices sites.
Ref. \cite{BEC} suggested that for non-interacting bosons ($U=0$) a
master equation (\ref{eq:masterequation}) with the above
Hamiltonian and quasi--local dissipation \begin{equation}
c_{\ell}\equiv
c_{ij}=\left(a_{i}^{\dagger}+a_{j}^{\dagger}\right)\left(a_{i}-a_{j}\right)\label{eq:jump}\end{equation}
has the unique steady state solution in form of a pure Bose
Einstein condensate representing a state with long range order \[
\left\vert \mathrm{BEC}\right\rangle =a_{\mathbf{q}=0}^{\dagger\,
N}/\sqrt{N!}\left\vert \mathrm{vac}\right\rangle .\] Here
$a_{\mathbf{q}}=1/\sqrt{M^{d}}\sum_{j}a_{j}e^{i\mathbf{q}\mathbf{e}_{\lambda}j}$
is the destruction operator for quasimomentum $\mathbf{q}$ on a
lattice with $M^{d}$ lattice sites, lattice vectors
$\mathbf{e}_{\lambda}$ and spacing $a$. It is easy to see that the
state $\left\vert \mathrm{BEC}\right\rangle $ satisfies
\begin{eqnarray*}
\textrm{(i)} & \forall\langle i,j\rangle\quad(a_{i}-a_{j})\left\vert \mathrm{BEC}\right\rangle =0\\
\textrm{(ii)} & H_{0}\left\vert \mathrm{BEC}\right\rangle
=N\epsilon_{\mathbf{q=0}}\left\vert \mathrm{BEC}\right\rangle
\end{eqnarray*} with $\epsilon_{\mathbf{q=0}}$ the single particle
Bloch energy for quasimomentum $\mathbf{q}=0$, and thus is
consistent with the conditions of a pure steady state of the master
equation. We show now that this state is unique.

Before proceeding we find it convenient to go to a quasimomentum
presentation. The dissipative part of the master equation then
takes on the form (\ref{eq:masterequation}) with jump operators
\begin{equation} \label{eq:jumpq} c_{\ell}\equiv
c_{\mathbf{q},\lambda}=\frac{1}{\sqrt{M^{d}}}\sum\limits
_{\textbf{k}}(1+\mathrm{e}^{\mathrm{i}(\textbf{k}-\mathbf{q})\textbf{e}_{\lambda}})(1-\mathrm{e}^{-\mathrm{i}\textbf{k}\textbf{e}_{\lambda}})a_{\textbf{k}-\mathbf{q}}^{\dagger}a_{\textbf{k}}\end{equation}
and couplings constants, $g_l$ independent of $l$.

To show uniqueness we show first off all that the BEC--state is the
unique dark state and that there exists no invariant subspace for
our choice of jump operators. For a fixed particle number $N$, the
first term in $c_\ell$ is a creation operator and has no
eigenvalues; in particular no zero eigenvalues. Thus, in order to
identify dark states $| D\rangle$ with zero eigenvalue, we may
restrict ourselves to the equation $(a_i - a_j)|D\rangle = 0
\,\,\forall\langle i,j \rangle$. Taking the Fourier transform, this
translates to $(1 - e^{i{\bf q} \textbf{e}_\lambda})a_{\bf q}
|D\rangle = 0 \,\,\forall\textbf{q}$. Thus the BEC state with
$\textbf{q} =0$ is the only dark state. Next, we will construct for
every state $|\Phi\rangle$ in the Hilbert space a polynomial
operator  $O(c_{{\bf q},\lambda})$, where the jump operators
$c_{{\bf q},\lambda}$ are given in Eq.~(\ref{eq:jumpq}), such that
$\bra{\mathrm{BEC}}O| \Phi\rangle\neq0$. With the notation ${\bf
n}=\{\ldots,n_{{\bf k}},\ldots\}$, the states
$\ket{\mathbf{n}}=\prod_{\mathbf{q}}(a_{\mathbf{q}}^{\dag})^{n_{\mathbf{q}}}\ket{0}$
form a basis in the Hilbert space, a general state can be written
as $\ket{\Phi}=\sum f_{\mathbf{n}}\ket{\mathbf{n}}$. We select a
state $\ket{\mathbf{m}}$ with $ f_{\mathbf{m}} \neq 0$ and the
number of particles in the zero momentum mode maximal. Furthermore,
for each ${\bf q}\neq 0$  we fix $\lambda_{\bf q}$ such that ${\bf
e}_{\lambda_{\bf q}}$ is not orthogonal on ${\bf q}$, i.e., ${\bf
e}_{\lambda_{\bf q}} {\bf q} \neq 0$. Note that $\langle
\mathrm{BEC}|(c_{{\bf q},\lambda_{\bf q}})^{n_{{\bf q}}}
(a^{\dag}_{{\bf k}})^{m_{{\bf k}}}|0\rangle \neq 0$ only if
$n_{{\bf q}}= m_{{\bf k}}$ and ${\bf q}={\bf k}$. Thus, applying
the polynomial operator $O=\Pi_{\mathbf{ q}\neq
0}\left(c_{\mathbf{q},\lambda_{{\bf q}}}\right)^{m_{\mathbf{q}}}$
to $\ket{\Phi}$ provides a finite overlap with the BEC, which
implies that the BEC is the unique stationary state of the
dissipative process.

\subsection{Preparation of Stabilizer States}
\label{subsec:drivenStab}

As a second example we show that stabilizer states can be obtained
as the unique stationary states of a dissipative process involving
only quasi--local interactions.

We denote by $X,Y,Z$ the standard Pauli operators. The Pauli group,
${\cal P}$, consists of all Pauli matrices, $X,Y,Z,\one$, together
with the multiplicative factors, $\pm 1, \pm i$. ${\cal P}_n={\cal
P}^{\otimes n}$ defines the Pauli group on $n$ qubits. A $n$--qubit
state, $\ket{\Psi}$, is called stabilizer state if it is uniquely
defined as the only eigenstate to eigenvalue $+1$ of a hermitian
subgroup (of order $n$) of ${\cal P}_n$, called the stabilizer of
$\ket{\Psi}$ \cite{Gothesis97}. That is, $\ket{\Psi}$ is the only
state left invariant under the subgroup

\bea S(\ket{\Psi})=\{g\in {\cal P}_n: g\ket{\Psi}=\ket{\Psi}\}.\eea

Let us denote the generators of this group by $U_l$, $l=1,\ldots
n$. A subset of the stabilizer states are the so--called Graph
states \cite{Hein}, which are associated to a mathematical graph
which consists out of vertices and edges. Whenever two vertices
$a,b$ are connected by an edge, we say that $b$ is in the
neighborhood of $a$ and write $b\in N_a$. A $n$--qubits Graph
state, $\ket{\Psi_{0,\ldots, 0}}$, can now be defined as the unique
eigenstate of a set of independent commuting observables $U_k= X_k
Z_{N_k}=X_k\prod_{b\in N_k} Z_b$, where $W_k=\sigma_w^k$, for
$W\in{X,Y,Z}$ denotes the Pauli operator $\sigma_w$ acting on qubit
$k$. Note that these unitaries define a unique basis, the
so--called Graph state basis, which we denote by
$\{\ket{\Psi_{i_1,\ldots ,i_n}}\}_{i_j\in \{0,1\}}$. It can be
shown that $\ket{\Psi_{i_1,\ldots ,i_n}}= Z_1^{i_1}\otimes \ldots
Z_n^{i_n} \ket{\Psi_{0,\ldots, 0}}$ \cite{Hein}, where $U_k
\ket{\Psi_{i_1,\ldots ,i_n}}=(-1)^{i_k}\ket{\Psi_{i_1,\ldots
,i_n}}$. In particular we have that $\ket{\Psi_{i_1,\ldots,
i_k=0,\ldots ,i_n}}= Z_k^{i_k} \ket{\Psi_{i_1,\ldots, i_k=1,\ldots
,i_n}}$. We consider now those Graph states for which all the
unitaries $U_k$ are quasi--local. An example would be the linear
Cluster state, where $U_k =Z_{k-1} X_k Z_{k+1}$. An other example
would be the 2D-Cluster state, which is a universal resource for
quantum computations \cite{RaBr01}. As shown in \cite{RaBr01}, once
such a state is prepared any quantum computation can be performed
by means of local measurements only.

We define the operators $c_k=\frac{1}{2}(\one+U_k)Z_k$, which act
only non--trivially on the same qubits as $U_k$ and are therefore
quasi--local. Note that $\one+U_k$ is a projector onto the subspace
of eigenvectors of $U_k$ with eigenvalue $+1$ and $Z_k$ changes any
eigenstate $\ket{\Psi_{i_1,\ldots, i_k=0,\ldots ,i_n}}$ to $
\ket{\Psi_{i_1,\ldots, i_k=1,\ldots ,i_n}}$ and visa versa. Thus,
$c_k$ is an operator which has only one $2^{n-1}$--fold degenerate
eigenvalue, namely $0$ and the corresponding eigenspace is
$\mbox{span} \{\ket{\Psi_{i_1,\ldots, i_k=0,\ldots ,i_n}}, \mbox{
with } i_j\in \{0,1\}, \mbox{ for } j\neq k\}$ Thus, we have that
$c_k \ket{\phi}=0$ iff $U_k \ket{\phi}=\ket{\phi}$, which implies
that the only state which is an eigenstate to all operators $c_k$
is the Graph state. Using the definition of $c_k$ it is easy to
show that $c_k=U\sigma_{-}^{(k)} U^\dagger$, where $U$ transforms
the computational basis into the Graph basis. Due to the results
presented in Sec. III, this shows that the Graph state is the
unique stationary state of the process described by ${\cal
L}_{\{c_k\}}$. Since any stabilizer state is up to some local
unitary (actually local Clifford) operation, $V$, equivalent to a
Graph state, the operators $c(S)_k=V c_k V^\dagger$ are
quasi--local and ${\cal L}_{\{c(S)_k\}}$ has as unique stationary
state the stabilizer state. Furthermore, the relaxation of this
process in independent of $n$ and there exists no purely imaginary
eigenvalue of the Liouvillian.

Note that we can write $c_{k}=Z_k P_k$, with
$P_k=\frac{1}{2}(\one-U_k)$ is the projector onto the eigenspace to
eigenvalue $-1$ of $U_k$. Thus, the evolution corresponding to
these operators can be implemented using a feedback mechanism.

Another way to show that the stabilizer states can be prepared
using quasi--local jump operators would be to use the relation
between the unitary which generates the state and the operators
$\{c_k\}$, as described in Section IV.

\subsection{Preparation of MPS--PEPS states}
\label{subsec:drivenMPS}

As mentioned above, one has to choose non--hermitian matrices to
guarantee that the desired state is the unique stationary state.
Since otherwise, at least the completely mixed state is a stationary
state too. The aim of this section is to demonstrate the general
construction of the jump operators. Therefore, we consider the
so--called product entangled pair state (PEPS)\cite{PVC06}. In one
dimension these states are called matrix product states (MPS). Let
us consider $n$ $d$--dimensional systems. Any state describing these
systems can be written as \bea \ket{\Psi}=\sum_{i_1,\ldots i_n=1}^d
\tr[A_{i_1}^{[1]}A_{i_2}^{[2]}\cdots A_{i_n}^{[n]}] \ket{i_1,\ldots
i_n},\eea where $A_{i_k}^{[k]}$ are $D\times D$ matrix, with $D$
being the bond dimension \cite{Vidal}. Using this way of presenting
the states and especially the generalization to 2D, has been proven
to be very powerful to determine, for instance, the ground states of
some Hamiltonians \cite{PVC06}.

Similar to the procedure above we design now the interaction
between the reservoir and the system, such that the desired
MPS--PEPS is the unique stationary state. In \cite{PVCM07} it has
been shown that for many MPS--PEPS one can construct a frustration
free Hamiltonian which has this state as the unique ground state.
These Hamiltonians are of the form \bea H=\sum_k h_k,\eea where the
hermitian operators $h_k$ are quasi--local. The ground state,
$\ket{\Psi}$, of these Hamiltonians is uniquely defined by the
equations \bea h_k \ket{\Psi}=\lambda_{\textrm{min}}\ket{\Psi}
\quad \forall k,\eea where $\lambda_{\textrm{min}}$ denotes the
minimal eigenvalue of $h_k$. That is, the ground state of the
Hamiltonian corresponds to the ground state of the quasi--local
Hamiltonians.

In order to obtain these states now as stationary states of a
dissipative process we construct the operators $c_k$ which have a
unique common eigenstate, which is the desired MPS--PEPS. In order
to do so we consider the ground states of the operators $h_k$,
$D(h_k)$ (note that this must be more than 1--dimensional) and
again construct the non--hermitian operators $c_k$ whose
eigenvectors span $D(h_k)$. Then, the only common eigenstate of the
operators $c_k$ is $\ket{\Psi}$. To ensure that there exists no
mixed stationary state, one might need to consider more than one
operator per site.

Let us illustrate the general idea by considering as an example the
ground state of the AKLT--model \cite{Aklt}. Historically, this
state occurred first in the context of condensed matter physics.
There it was shown to be the unique ground state of a
Heisenberg--like Hamiltonian (see below) \cite{Aklt}. Recently this
state has attracted interest in quantum information theory, due to
its useful properties for quantum communication \cite{VeDe}. The
Hamiltonian has the following form: \bea H=\sum_k h_k, \quad
\mbox{with } h_k=\vec{S}_k \vec{S}_{k+1} +\frac{1}{3}
\left(\vec{S}_k \vec{S}_{k+1}\right)^2,\eea where the operators
$S^{\alpha}$ with $\alpha\in\{x,y,z\}$ denote the spin-1 operators.
The quasi--local Hamiltonians, $h_k$ act non--trivially on system
$k$ and $k+1$. As mentioned before, the ground state of the
Hamiltonian, which we denote by $\ket{\Psi}$, is the ground state
of all quasi--local Hamiltonians, $h_k$.

Since the Hamiltonian is translational invariant we omit the index
$k$ as well as the identity operator whenever it does not lead to
any confusion. The operators $h_k$ have the eigenvalues $4/3$,
which is $5$--fold degenerate and $-2/3$, which is four--fold
degenerate. We denote by $\ket{\Phi_k}$ ($\ket{\Psi_k}$) an
orthonormal basis of the subspace corresponding to the eigenvalue
$4/3$ ($-2/3$) respectively. The operators $c_k$ (acting on system
$k$ and $k+1$) can be defined as: \bea \label{eq_ci}
c_k&=&\ket{\Psi_1}\bra{\Phi_1}+\ket{\Psi_2}\bra{\Phi_2}+
\ket{\Psi_3}\bra{\Phi_3}\\
\nonumber &+&\ket{\Psi_4}\bra{\Phi_4}+\ket{\Phi_4}\bra{\Phi_5}.\eea

It is easy to see that the only states which are eigenstates of
$c_k$ are states in the subspace spanned by
($\{\ket{\Psi_k}\}_{k=1}^4$), i.e. the ground state subspace of
$h_k$. Thus, the set of operators $\{c_k\}$ is sufficient to
guarantee that there is no other pure stationary state.

In order to show that there is no mixed state, we note that the
operators $c_k$ can be written as $c_k=U_k P_k$, where $U_k$ is a
unitary matrix and $P_k=\frac{1}{2} (2/3 \one+ h_k)$ is the
projector onto the eigenstates of $h_k$ with eigenvalue $4/3$. The
choice of the unitary is by no means unique and we will use this
freedom to show that also in this case there is no mixed stationary
state. Let us write $c_k^\alpha=U_k^\alpha P_{k}$, then
$c_k^\alpha\ket{\Phi}=0$ $\forall k, \alpha$ iff
$\ket{\Phi}=\ket{\Psi}$. For any state $\ket{\chi}$ different than
$\ket{\Psi}$, there exists a $k$ such that $P_k\ket{\chi}\neq 0$.
Without loss of generality we assume $k=n$. We construct now a
completely positive map, ${\cal E}_k(\rho)=\sum_\alpha
(c_k^\alpha)^\dagger \rho c_k^\alpha$ such that ${\cal
E}(\proj{\Psi})\equiv {\cal E}_n \circ{\cal E}_{n-1}\circ \ldots
\circ{\cal E}_1(\proj{\Psi})\propto \one_{2,3,\ldots ,n}\otimes
P_{n}$ and therefore, $\bra{\chi}{\cal E}(\proj{\Psi}) \ket{\chi}
\neq 0$. Since all the terms occurring in ${\cal E}(\proj{\Psi})$
are positive semidefinite, this implies that there exists a
monomial $P=c_n^{\alpha_n}\ldots c_1^{\alpha_1}$ such that
$\bra{\chi}P^\dagger \proj{\Psi}P \ket{\chi}=| \bra{\Psi}P
\ket{\chi}|^2  \neq 0$. Thus, the state is unique. We choose
sufficiently many $U^\alpha$ such that $\sum_\alpha
(U_k^\alpha)^\dagger \rho U_k^\alpha =\frac{1}{9}\one_k\otimes
\tr_{k,k+1} (\rho)$ for any state $\rho$ \footnote{ This can always
be achieved with finitely many $U_k^\alpha$ since the dimension of
the Hilbert space is finite, see for instance \cite{DuHe05}.}.
Using that $\tr_k(P_k)\propto \one$, we find \bea {\cal E}_{2}
({\cal E}_{1}(\proj{\Psi})\propto {\cal E}_{2} (P_1\otimes
\tr_{1,2}(\proj{\Psi})\propto
\\ \nonumber
\one_1\otimes P_2 \otimes \tr_{1,2,3}(\proj{\Psi}).\eea Continuing
in this way we end up with $\one_{2,3,\ldots ,n}\otimes P_{n}$,
which shows that the ground state of the AKLT--model is unique.

\subsection{The driven $\eta$-condensate}
\label{subsec:drivenEta} The second example for a dissipatively
driven state given in~\cite{BEC} is the $\eta$-condensate of paired
fermions. First described by Yang~\cite{Yang89}, the
$\eta$-condensate is an exact excited eigenstate of the
Fermi-Hubbard (FH) model for fermions with two internal states,
$\uparrow,$ $\downarrow$, on a bipartite lattice in $d$ dimensions
with $M^d$ sites. Its properties derive from the fact that the
FH-Hamiltonian
\begin{equation}\label{HFH}H_{FH}=-J\sum_{\langle l,l' \rangle,\sigma=\uparrow,\downarrow}f^{\dagger}_{\sigma,l}
f_{\sigma,l'}+U\sum_l
f^{\dagger}_{\uparrow,l}f_{\uparrow,l}f^{\dagger}_{\downarrow,l}
f_{\downarrow,l}\end{equation} and the $\eta$-creation operator
\begin{equation}\label{etaop}\eta^{\dagger}\equiv \frac{1}{\sqrt{M^d}}\sum_l S(l)
f^{\dagger}_{\uparrow,l}f^{\dagger}_{\downarrow,l}\end{equation}
obey the commutation relation
$[\eta^{\dagger},H_{FH}]=U\eta{\dagger}$, where $S(l)$ in
(\ref{etaop}) alternates between $\pm1$ in a checkerboard pattern.
The $f^{\dagger}_{\sigma,l}$ operators create a fermion with spin
$\sigma=\uparrow,$ $\downarrow$ on site $l$ and obey canonical
anticommutation relations. In (\ref{HFH}), $J$ denotes the tunneling
rate and $U$ the on-site inter-species interaction.

From this observation, it follows that the state
$|\eta,N\rangle\equiv\left(\eta^{\dagger}\right)^N|0\rangle$ is an
exact excited eigenstate of $H_{FH}$, with energy $NU$, irrespective
of the boundary conditions. In the following, we assume periodic
boundary conditions. In position space the $\eta$-condensate can be
understood as a superposition state of all possible vectors
$|D^{\{d_k\}}\rangle$, which denote position basis states in
$\mathcal{H}_N$ that have only double occupation, at positions
$d_k$, $k\in[1,\ldots,N]$:
$|\eta,N\rangle=\sum_{\{d_k\}}S(\{d_k\})|D^{\{d_k\}}\rangle$, where
$S(\{d_k\})=\pm1$ if the number of even $d_k$ is itself even (odd)
on the even-indexed sites. A pair of fermions on the same site is
called a \emph{doublon} in the following.

The $\eta$-condensate, while never a ground state of the
Hubbard-model, is interesting from the perspective of many-body
physics, as it exhibits perfect superfluidity in any spatial
dimension, without any approximations. $\eta$-pairing
($\langle\eta^{\dagger}\eta\rangle\neq0$) in the ground state has
also been considered for doped negative $U$ Hubbard
models~\cite{Singh91} and extended Hubbard models (see e.g.
~\cite{Essler92}).

We define the jump operators
\begin{eqnarray}c^{(1)}_l&=&(\eta^{\dagger}_{l}-\eta^{\dagger}_{l+1})(\eta_{l}+\eta_{l+1})\\
c^{(2)}_l&=&n_{\uparrow,l}f^{\dagger}_{\downarrow,l}f_{\downarrow,l+1}+n_{\uparrow,l+1}f^{\dagger}_{\downarrow,l+1}f_{\downarrow,l}\nonumber\\
c^{(3)}_l&=&(f^{\dagger}_{\uparrow,l}+f^{\dagger}_{\uparrow,l+1})(f_{\uparrow,l}+f_{\uparrow,l+1})(1-n_{\downarrow,l})(1-n_{\downarrow,l+1})\nonumber\\
c^{(4)}_l&=&(f^{\dagger}_{\downarrow,l}f_{\downarrow,l+1}+f^{\dagger}_{\downarrow,l+1}f_{\downarrow,l})n_{\uparrow,l}n_{\uparrow,l+1}
\nonumber\end{eqnarray} on the Hilbert space $\mathcal{H}_N$, in
which all states have $N$ spin-up and $N$ spin-down fermions. Here,
$\eta^{\dagger}_{l}\equiv
f^{\dagger}_{\uparrow,l}f^{\dagger}_{\downarrow,l}$, and
$n_{\sigma,l}\equiv f^{\dagger}_{\sigma,l}f_{\sigma,l}$.

A dark state $|\Psi\rangle\in\mathcal{H}_N$ is defined through the
condition $c^{(k)}_l|\Psi\rangle=0$ $\forall$ $k$, $l$. In the
orthonormal position basis, $|\Psi\rangle$ is defined through
coefficients $A_{i_1,j_1;\ldots;i_M,j_M}$, where $i_l=1$ ($=0$) if
there is (no) spin-up fermion on $l$, and $j_l$ being analogously
defined for spin-down fermions. Besides normalization, all $A$'s
must obey $\sum_{l=1}^{M^d} i_l=\sum_{l=1}^{M^d} j_l=N$. In this
notation, $c^{(1)}_l$ imposes that the coefficients for any dark
state must obey $A_{1,1;0,0}=-A_{0,0;1,1}$ and
$A_{1,1;1,0}=A_{1,1;0,1}=A_{1,0;1,1}=A_{0,1;1,1}=0$ $\forall$
$i_k$, $j_k$, $k\neq l,l+1$ (the $i_k$'s and $j_k$'s in $A$ are
suppressed for brevity). The first of the two conditions is the
essential property of the $\eta$-condensate: $|D^{\{d_k\}}\rangle$
differing in the position of one doublon by one site carry opposite
sign. $c^{(2)}_l$ imposes the constraint
$A_{1,0;0,1}=A_{0,1;1,0}=0$, which signifies that fermions of
opposite spin on adjacent sites are associated into doublons (on
the site of the spin-up fermion). $c^{(3)}_l$ then imposes
$A_{1,0;0,0}=-A_{0,0;1,0}$ and $A_{1,0;1,0}=0$. These jump
operators induce a diffusion process, which delocalises the spin-up
fermions provided they do not encounter spin-down fermions or
doublons. Furthermore two spin-up fermions may not sit on adjacent
sites in these dark states. $c^{(4)}_l$ finally imposes
$A_{1,1;1,0}=-A_{1,0;1,1}$, i.e. spin-down fermions must also be
delocalised over the lattice, but only atop the spin-up fermions.

Using the conditions imposed by $c^{(1)}$, $c^{(2)}$ and $c^{(3)}$
and the constraint on the number of spin-up and spin-down fermions,
it is straightforward to show that the $\eta$-condensate is the only
dark state. Note that $|\eta,N\rangle$ is trivially also a dark
state to $c^{(4)}$. We assume that there are nonzero coefficients in
$|\Psi\rangle$ for configurations with unpaired spin-up and
spin-down fermions. The conditions for the dark state then
immediately yield a contradiction. This is the case, as the
condition $A_{1,0;0,0}=-A_{0,0;1,0}$ means that any unpaired spin-up
fermion may be shifted in position if no spin-up fermion, spin-down
fermion or doublon is on the adjacent site. By assumption and the
condition, all the coefficients to these configurations must also be
nonzero. Shifting sufficiently often, one of these obstacles will
eventually be encountered - at which point the other conditions will
yield a contradiction.

Thus, any dark state may only have nonzero coefficients for the
doublon basis states $|D^{\{d_k\}}\rangle$. The first condition from
the $c^{(1)}$ then immediately yields that all these coefficients
must be equal to $S(\{d_k\})$. This is the case, because starting
from any doublon configuration with nonzero amplitude, repeated
application of this condition yields that any other possible doublon
configuration on the lattice must also be nonzero, with its sign
also obeying the rules for the $\eta$-condensate.

The proof of uniqueness again requires that for any
$|\Psi\rangle\in\mathcal{H}_N$ a monomial $P(\{c^{(k)}_l\})$,
$k=1,\ldots,4$, can be constructed s.t.
$\langle\eta,N|P|\Psi\rangle\neq0$. The proof proceeds differently
for two different cases: a) For $|\Psi\rangle\notin
\mbox{span}\{|D^{\{d_k\}}\rangle\}$, where $|D^{\{d_k\}}\rangle$
denotes a basis state carrying only doublons on sites $d_k$,
$k\in[1,\ldots,N]$. b) For $|\Psi\rangle\in
\mbox{span}\{|D^{d_k}\rangle\}$.

a) The proof for the uniqueness of the dark state indicates that we
can proceed analogous to the case of the BEC, but now in position
space instead of momentum space: Pick one position basis state
$|\phi^{(0)}\rangle$ with unpaired fermions occurring in
$|\Psi\rangle$ with nonzero amplitude. It is clear from the above
proof that we can always find a sequence of applications of
$c^{(3)}_l$'s and $c^{(4)}_l$'s s.t. a spin-up and spin-down fermion
are on adjacent site (we observe that $c^{(4)}_l$ allows a doublon
to swap positions with a spin-up fermion on an adjacent site).
Application of $c^{(2)}$ then associates these into a doublon. It is
straightforward to see that this sequence can be constructed s.t. it
maps $|\phi^{(0)}\rangle$ into a particular $|D^{\{d_k\}}\rangle$,
and any state orthogonal to $|\phi^{(0)}\rangle$ to zero.

b) We can proceed directly analogous to the proof of uniqueness for
the ground state of the AKLT-model, by generalizing the jump
operators $c^{(1)}_l$ to $c_l^{(1)\alpha}=U^{\alpha}_lP_l$, with
$P_l$ the projector on the symmetric state of one doublon on sites
$l$ and $l+1$. The unitaries, $U^{\alpha}_l$ now must conserve
particle number. We define completely positive maps $\mathcal{E}_l$
again as in \ref{subsec:drivenMPS}. Applying them in sequence on
the $\eta$-condensate yields a density matrix with nonvanishing
overlap to $|\Psi\rangle$, which proves the existence of a monomial
$P$ leading to finite overlap, as for the ground state of the
AKLT-model.

\section{Conclusion and Outlook}

In this paper we have shown how to use dissipative processes for
state preparation. That dissipation, in conjunction with
measurements, can be used to prepare pure states is known (see e.g.
\cite{BarBou02,Roa06}), where usually single- or few-particle
states are considered. Here, we demonstrated how dissipation can be
employed to generate multipartite states. For an arbitrary $n$
$d$--level state we constructed a dissipative process (with
constant relaxation time), consisting of $n$ jump operators, which
has this state as the unique stationary state. For certain states,
like the Cluster states, we showed that this process can be
implemented using only quasi--local operations. Apart from that we
demonstrated how a quasi--local dissipative process, which is
suitable for state preparation, can be constructed for a given
state, which has a quasi--local description. We illustrated this
method considering the ground state of the AKLT--model.
Furthermore, we showed that the quasi--local dissipative processes,
which we considered in \cite{BEC}, have as unique stationary states
the BEC--state, and the $\eta$--condensate, respectively.

The processes discussed here might be also used to gain some
insight in the computational complexity of certain problems, like
for instance the so--called satisfiability problem \cite{Kitaev02}.
There, the aim is to find out whether there exists a common
solution to a set of Boolean equations. To be more precise, the
problem is to determine a $n$--bit string for which
$m=\mbox{poly}(n)$ given clauses, involving only 3 (2) variables
for the 3-SAT (2--SAT) problem respectively, hold true. It is
straightforward to construct a dissipative process such that any
computational state, corresponding to the classical solution of the
problem, is a stationary state of the process. Using the formalism
we developed here, we are going to investigate the difference
between the NP--complete $3$--SAT problem and the $2$--SAT problem,
which can be solved (classically) in polynomial time. It might well
be that looking at this problem from this completely different
point of view leads to a new insight. Furthermore, it would be
interesting to establish a relationship between the relaxation time
and the non--locality of the jump operators, which cause the system
to evolve to the desired state. Apart form that, the investigation
of dissipative processes, which lead to higher dimensional dark
state subspaces might lead to interesting applications for quantum
computation and storage of quantum information.

In \cite{VeWo08} the authors investigated independently similar
aspects of dissipative processes. It has been shown there, that
these processes can also be employed for universal efficient
quantum computation.

\section*{Acknowledgements}

B.K. would like to thank M. Lewenstein, M. van den Nest and E. Rico
for helpful discussions. We acknowledge support by the Austrian
Science Foundation (FWF), the European Commission through the
Integrated Project FET/QIPC SCALA, as well as OLAQUI and QICS, and
the Elise Richter Program.

\end{document}